\documentclass[lettersize,journal]{IEEEtran}
\IEEEoverridecommandlockouts
\usepackage{etoolbox}
\makeatletter
\patchcmd{\@makecaption}
  {\scshape}
  {}
  {}
  {}
\makeatletter
\patchcmd{\@makecaption}
  {\\}
  {.\ }
  {}
  {}
\makeatother
\def\tablename{Table}

\usepackage{cite}
\usepackage{amsmath,amssymb,amsfonts}
\usepackage{algorithmic}
\usepackage{graphicx}
\usepackage{textcomp}
\usepackage{xcolor}
\usepackage{booktabs}

\usepackage{lettrine}

\usepackage{titlesec}
\usepackage{times}
\usepackage{enumitem}
\usepackage{multirow}
\usepackage{caption} 
\usepackage{subcaption}
\usepackage{microtype}
\usepackage{balance}
\def\BibTeX{{\rm B\kern-.05em{\sc i\kern-.025em b}\kern-.08em
    T\kern-.1667em\lower.7ex\hbox{E}\kern-.125emX}}

\titlespacing*{\section}
{0pt}{1.5ex plus 1ex minus .2ex}{0.3ex plus .2ex}
\titlespacing*{\subsection}
{0pt}{1.0ex plus 1ex minus .2ex}{0.3ex plus .2ex}
\titlespacing*{\subsubsection}
{0pt}{0.8ex plus 0.5ex minus .2ex}{1ex}

\providecommand{\tabularnewline}{\\}
\newcommand{\mwu}[1]{#1}

\begin{document}

\title{StruM: Structured Mixed Precision for Efficient Deep Learning Hardware Codesign}

 \author{
     \IEEEauthorblockN{Michael Wu, Arnab Raha, 
     Deepak A. Mathaikutty, Martin Langhammer, Engin Tunali, and Daksha Sharma\\}
     \IEEEauthorblockA{Intel Corporation, CA, USA.}
     }

\maketitle
\thispagestyle{plain} 
\pagestyle{plain}

\begin{abstract}
In this paper, we propose StruM, a novel structured mixed-precision-based deep learning inference method, co-designed with its associated hardware accelerator (DPU), to address the escalating computational and memory demands of deep learning workloads in data centers and edge applications. Diverging from traditional approaches, our method avoids time-consuming re-training/fine-tuning and specialized hardware access. By leveraging the variance in weight magnitudes within layers, we quantize values within blocks to two different levels, achieving up to a 50\% reduction in precision for 8-bit integer weights to 4-bit values across various Convolutional Neural Networks (CNNs) with negligible loss in inference accuracy. To demonstrate efficiency gains by utilizing mixed precision, we implement StruM on top of our in-house FlexNN DNN accelerator~\cite{flexnn} that supports low and mixed-precision execution. Experimental results depict that the proposed StruM-based hardware architecture achieves a 31-34\% reduction in processing element (PE) power consumption and a 10\% reduction in area at the accelerator level. In addition, the statically configured StruM results in 23-26\% area reduction at the PE level and 2-3\% area savings at the DPU level.
\end{abstract}


\section{Introduction}\label{sec_Introduction}
\noindent Deep learning workloads have gained immense significance in both datacenter and edge applications, posing a significant challenge due to the substantial surge in computational and memory bandwidth demands. These workloads entail complex operations, including many convolution operations and large matrix multiplications, executed on large datasets.  As deep learning techniques advance, there is a continuous drive to improve the accuracy of models. However, these improvements lead to significant expansions in both model parameter sizes and operation counts. 
In response to these escalating computational demands, strategies have been developed to mitigate the burden. One approach focuses on crafting efficient deep learning network architectures~\cite{szegedy2015going, iandola2016squeezenet, howard2017mobilenets,tan2021efficientnetv2, zhang2018shufflenet,chollet2017xception, zoph2017neural, cai2018proxylessnas,tan2019mnasnet}. Additionally, another approach concentrates on reducing the computational cost of convolution and matrix multiplication operations. This includes quantization methods that perform \textit{Multiply-Accumulate~(MAC)} operations at lower precision~\cite{zhou2016dorefa,hubara2016binarized, umuroglu2017finn,gong2018highly, jain2019trained, zhu2020xor}, as well as pruning methods that aim to skip MAC operations by leveraging zero values in weights. 

\begin{figure*}[h]
\centering
\includegraphics[width=0.98\textwidth]{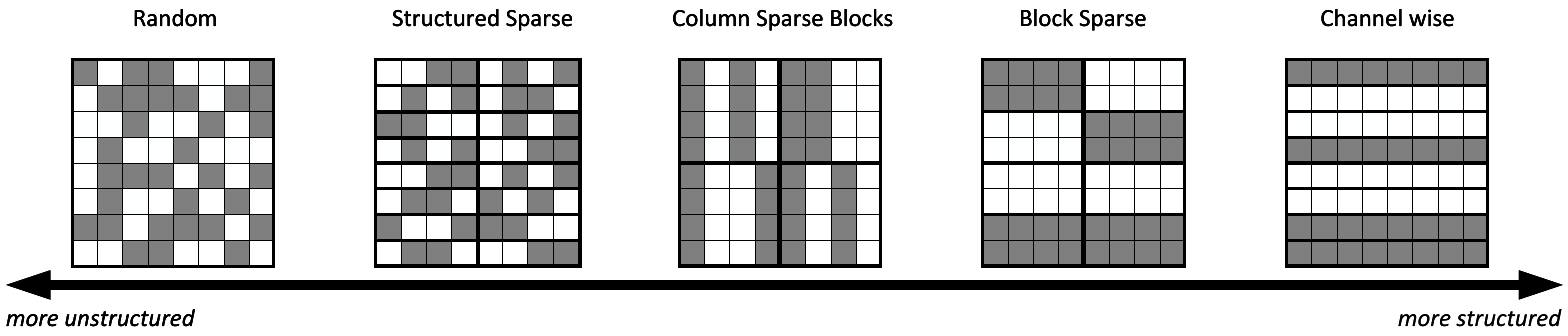}
\captionsetup{belowskip=0pt}
\caption{An overview of different pruning methods. Gray squares are nonzero values while white squares are zero values. }
\label{fig:pruning}
\end{figure*}

Neural network pruning methods aim to reduce computational load by eliminating low-impact weights, typically by setting weights with small magnitudes to zero. Fig.~\ref{fig:pruning} provides an overview of various pruning methods, such as unstructured sparsity~(or random sparsity), channel-wise pruning, and block-wise pruning. Unstructured sparsity~\cite{han2015deep,han2015learning} allows random placement of zeros, while channel-wise~\cite{li2016pruning}, column-wise, and block-wise pruning)~\cite{narang2017block,gray2017gpu} induce more structured sparsity, optimizing for hardware efficiency. NVIDIA employs structured sparsity~\cite{mishra2021accelerating}, an intermediate method between completely unstructured and channel-wise pruning, where a fixed number of zeros are induced in each block but can be at any location within the block, enabling computational and hardware complexity savings. However, highly structured pruning methods sacrifice classification accuracy compared to less structured approaches. Therefore, retraining or fine-tuning is essential to restore model performance, particularly for low bitwidth and/or very sparse weights~\cite{han2015deep, han2015learning,li2016pruning,umuroglu2017finn,jain2019trained,gong2018highly,
zhou2016dorefa,hubara2016binarized,zhu2020xor}. \mwu{Unfortunately, the typical retraining/finetuning process faces several challenges related to speed and computational demands:
\begin{enumerate}[label=\roman*.]
\item A software infrastructure is needed for structured sparsity retraining/finetuning.
\item The retraining/finetuning process can be time-consuming, especially for large-scale language models~(LLMs). For example, training state-of-the-art LLMs can require millions of dollars in computational resources~\cite{cottier2024rising}.
\item Hyperparameter tuning is often necessary to achieve satisfactory accuracy and acceptable convergence speed.
\end{enumerate}
In addition, there are also challenges concerning the division of model and implementation responsibilities:
\begin{enumerate}[label=\roman*.]
\item Customers, such as Microsoft or Meta, typically provide a model to vendors such as AMD, Intel, or Qualcomm.
\item Vendors are required to quantize the model to enable efficient inference, but the customer does not tune their model for specific hardware.
\item Datasets are often unavailable to the end customer, preventing them from performing effective retraining or finetuning.
\end{enumerate}}

In this paper, we introduce \textit{Structured Mixed Precision~(StruM)}: a method incorporating regular structures in weights, combining low and high precision weights. This approach facilitates efficient hardware implementations and maintains high network accuracy, \mwu{\textbf{eliminating the need for retraining/finetuning}}. To showcase the hardware efficiency, we modified our in-house \textit{FlexNN} accelerator microarchitecture~\cite{flexnn} to incorporate StruM. The flexibility in FlexNN architecture stems from its ability to support multiple dataflows that can be configured to support the optimal dataflow for each layer of a DNN. FlexNN also features two-sided unstructured sparsity acceleration while supporting various activation and weight datatypes such as INT8, FP16, \emph{etc.} 

In a nutshell, StruM's key contributions include:
\begin{enumerate}
\item StruM introduces structured mixed precision for hardware acceleration, partitioning weights into blocks and employing different quantization strategies for each set within the block. This approach optimizes weight partitioning for hardware efficiency and introduces two quantization strategies, DLIQ and MIP2Q, to reduce hardware complexity while maintaining classification accuracy. 
\item Structured mixed precision also alleviates the impact of the slowest PE effect (similar to structured sparsity) by balancing low precision operands across PEs ensuring close to optimal low precision acceleration at the DNN accelerator level resulting in close to ideal performance for a target precision ratio.
\item Both DLIQ and MIP2Q methods exhibit minimal classification accuracy loss (less than $<1\%$) compared to the INT8 baseline across diverse neural network architectures for ImageNet classification. They consistently surpass structured sparsity and maintain stable accuracy across different parameter settings, highlighting StruM's potential for hardware implementation.
\item We implemented StruM on top of our in-house flexible dataflow DNN accelerator known as FlexNN~\cite{flexnn} that supports multi-precision convolution and matmul operation with two-sided unstructured sparsity acceleration. 
\item Strum-based processing elements demonstrate a 31-34\% reduction in processing element (PE) power consumption and a 10\% reduction in power consumption at the accelerator level. In addition, the statically configured StruM results in 23-26\% area reduction at the PE level and 2-3\% area savings at the accelerator level. For the dynamically configured StruM, we see a 3\% additional area overhead at the accelerator level with the same power savings.
\end{enumerate}

The rest of the paper is organized as follows. Section~\ref{sec_Related_Work} provides a brief background on the pros and cons of unstructured \emph{vs.} structured sparsity that extends to mixed precision as well. Section~\ref{sec_Motivation} describes the background and motivation behind StruM. This is followed by Section~\ref{sec_design} which describes the StruM technique and how it is used to train the DNN model. Subsequently, Section~\ref{sec_flexnn} briefly introduces the microarchitecture of our in-house DNN accelerator called FlexNN followed by a detailed description of how StruM (or structured mixed precision) was implemented on top of the existing FlexNN design. This is followed by Section~\ref{sec_exp_meth} which explains the software and hardware experimental methodology. Next, Section~\ref{sec_Results_and_Discussions} provides a detailed account of experimental results mostly highlighting the area and power savings considering the FlexNN accelerator as the baseline. It also reports the accuracy impact of StruM on the application-level output quality. This section also discussed some of our assumptions and potential future optimizations. Finally, Section~\ref{sec_Conclusion_and_Future_Work} concludes the paper.

\section{Related Work}\label{sec_Related_Work}
In this section, we explore the existing literature relevant to our proposed method, considering various techniques. In the case of unstructured sparsity~\cite{han2015deep,han2015learning}, there are no constraints on the locations of the zeros. The generation of zero-value weights during pruning typically involves setting weights with small absolute magnitudes to zero. To increase the number of weights that can be set to zero, retraining/fine tuning is necessary. In~\cite{han2015deep}, an iterative process is introduced. After pruning, the network undergoes fine-tuning to recover lost accuracy. The process is then iterated, with more and more connections pruned in each step to achieve higher sparsification rates. In \cite{han2015learning}, iterative pruning and retraining are compared against one-shot retraining, with iterative pruning shown to be more effective compared to the one-shot method. Nevertheless, retraining/fine tuning is essential to achieve high compression rates for unstructured sparsity.

The main challenge with unstructured sparsity is that, due to its finer-grained data access patterns, it does not align well with typical ML accelerators, which often employ dense matrix multiplication units or vector units to speed up machine learning operations~\cite{jia2022xvdpu,flexnn,choquette2023nvidia}. \mwu{Consequently, unstructured sparse techniques do not effectively leverage hardware accelerators. To better utilize hardware resources like multiplication units, it's beneficial to induce regular sparse structures during the pruning process}. One common method for inducing regular structures is through channel-wise pruning~\cite{li2016pruning,luo2017thinet}. This technique involves removing entire filter channels from convolutional layers.  Channels with low importance, for example, identified by small L1 norms, are removed. Eliminating these channels results in a structured sparsity pattern that can be efficiently processed by matrix multiplication and/or vector units. However, after channel-wise pruning, it's crucial to fine-tune or retrain the pruned network. This step is necessary to mitigate performance degradation resulting from channel-wise pruning's inherently coarse nature compared to unstructured pruning~\cite{siswanto2021reconciling}.

Alternatively, another method for structured pruning involves removing weights block-wise~\cite{narang2017block}. In contrast to channel-wise pruning, which operates at a coarser level by removing entire filter channels, block-wise pruning offers a finer approach by targeting small blocks of channel weights. Similar to unstructured sparsity and channel-wise pruning, block-wise pruning identifies and removes low-importance weights, which can be characterized by small L1-norm values. To support the efficient processing of networks with block sparse weights, highly optimized GPU kernels have been specifically designed~\cite{gray2017gpu}. These kernels are tailored to handle the computational patterns arising from block-wise sparsity. However, to maintain classification accuracy after pruning, retraining or fine-tuning of the pruned network is essential for this method as well.

NVIDIA utilizes a finer-grained pruning technique known as structured sparsity in their hardware architecture, such as H100/A100. For their implementation, for every four weights in a channel, two of them are set to zero or pruned, resulting in a sparsity ratio of 50\%. NVIDIA's GPUs are optimized to effectively execute operations on tensors with this 2:4 sparsity pattern, enhancing overall computational efficiency during neural network inference~\cite{mishra2021accelerating,choquette2023nvidia}. However, due to the small block size and the fixed sparsity ratio of 50\%, retraining is necessary to recover classification accuracy. Without retraining, there is a significant decline in classification accuracy~\cite{pool2020accelerating}.

Another way to reduce hardware complexity is by employing quantization, which enables the use of lower precision multipliers like INT8 or INT4. Quantization methods are often combined with pruning techniques, where models undergo quantization followed by pruning, or both quantization and pruning are jointly considered~\cite{tung2018deep}. An alternative approach is to perform convolutions in the log domain, as suggested in~\cite{lee2017lognet,elhoushi2021deepshift}. In the log domain, multiplications are transformed into shifts and additions, significantly reducing computational complexity. However, this method comes with a significant drawback of decreased classification accuracy, necessitating retraining.

\mwu{Finally, there are methods that use mixed precision, assigning different precisions to weights and activations (full or partial) to accelerate inference. In Dynamic Region-based Quantization (DRQ)~\cite{DRQ}, varying levels of quantization are applied to feature map values to reduce computational complexity. First, feature map into rectangular regions and a threshold value is used to partition the regions into sensitive regions and insensitive regions. Higher precision (\emph{e.g.,} int8xint8) is assigned to sensitive regions, while insensitive regions use lower precision (\emph{e.g.,} int4xint4). In contrast, Hardware-aware Automated Quantization with Mixed Precision (HAQ)~\cite{HAQ} employs a reinforcement learning (RL) framework to automate the selection of optimal bitwidths for weights and activations on a per-layer basis across various hardware accelerators. Like before, the common drawback of both methods is that retraining or fine tuning is necessary to achieve good classification accuracy.}

In our proposal, we introduce a technique aimed at creating a hardware-friendly architecture without necessitating retraining or fine-tuning to maintain high classification accuracy.
%

\section{Motivation and Background}\label{sec_Motivation}
The motivation behind structured mixed precision stems from the performance loss that arises from unstructured low or mixed precision (or sparsity) support. Two prominent ways of introducing mixed low precision in a DNN model exist today. The most popular way to train a mixed precision DNN model is to train the model where either all the layers (mostly except the first and/or the last layers of the model) are trained uniformly to a low precision type. An alternate way is to train the DNN model layer-wise with different mixed precision types. In the latter case, each layer can have its precision type. Prior art belonging to these types have been discussed in detail in Section~\ref{sec_Related_Work}.


In contrast to the two existing ways of introducing mixed precision in DNN models, we explore the possibility of incorporating mixed precision weights at a finer granularity. Each layer can have multi-precision weights but they can occur randomly at any channel index. Not only the lower precision weights can be exploited to reduce the memory storage requirement and the bandwidth requirement but they can also be leveraged to reduce the power consumption of the MAC operation. However, this can only be achieved if the MAC unit is designed to execute the low precision weights with lower complexity leading to lower power consumption.

Note that an accelerator can also be designed to exploit lower precisions for performance acceleration as we can process a larger number of operands per clock cycle. Towards that end, StruM can potentially alleviate the impact of the slowest PE effect (similar to structured sparsity) by balancing low precision operands (in this case weights) across PEs ensuring close to optimal low precision acceleration at the DNN accelerator level resulting in close to ideal performance for a specified precision ratio (similar to sparsity percentage). Although we have not enabled this feature explicitly for our case, we have mentioned how it can be enabled without too much overhead in FlexNN in the Section~\ref{sec_design}.



\section{Structured Mixed Precision}\label{sec_design}
\noindent In this section, we start with a general high-level overview of StruM. Following that, we break down each step of StruM, explaining how it contributes to hardware acceleration. 
\subsection{General Description}
\label{sec:general_description}
For structured mixed precision, we partition weights into blocks, with each set utilizing a distinct quantization method/configuration. One set employs higher precision quantization, while the other set uses lower precision quantization. Specifically, a fixed number of values within each block are assigned to high precision, and another fixed number of values are assigned to low precision. The process involves the following steps:

\begin{itemize}
\item \textbf{Block Division:} We divide the weights into blocks of size $[l,w]$, where both $l$ and $w$ are positive integers. The specific method of partitioning is hardware-dependent, as no single approach suits all hardware configurations. A tailored partitioning strategy for our hardware will be detailed in Sec.~\ref{subsec:part}.

\item \textbf{Set Quantization:}\label{sec:step_2} Within each block, the values are partitioned into two independent sets, each employing a different quantization strategy. One set, with $p$ percent of the values, is quantized to lower precision, while the other set, with $1-p$ percent of the values, is quantized to higher precision. In Sec.~\ref{subsec:setq}, we will outline various set quantization methods that achieve different hardware complexity and classification accuracy tradeoffs.

\item \textbf{Runtime Processing:} At runtime, the hardware processes the two sets at different precisions to optimize computational efficiency--more resources are allocated to the higher precision set,  while fewer resources are dedicated to the lower precision set.
\end{itemize}
\subsection{Hardware Aware Block Division}\label{subsec:part}
First, we detail the partitioning of weights into blocks, optimizing for efficient processing on typical ML accelerators.\footnote{The proposed partition method is one of several feasible approaches.} 
In \textit{Convolutional Neural Networks (CNNs)}, weights are represented as 4-dimensional tensors $(f_h,f_w,f_d,f_c)$, where $f_h$ denotes height, $f_w$ represents width, $f_d$ indicates depth (or the number of input channels), and $f_c$ signifies the number of output channels. Considering that each output channel can be processed independently, we focus on one output channel without loss of generality.
%
\begin{figure}[h]
\centering
\includegraphics[width=0.95\columnwidth]{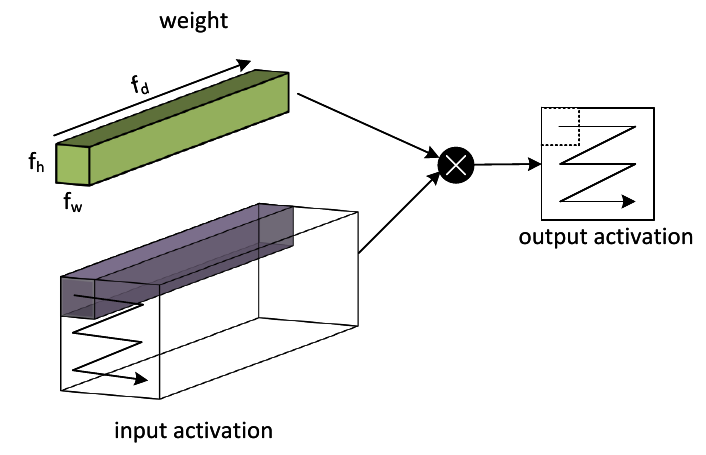}
\captionsetup{belowskip=0pt}
\caption{An example of a $1\times1$ convolution}
\label{fig:conv}
\centering
\end{figure}

Consider the $1\times 1$ convolution filter depicted in Fig. \ref{fig:conv}, where $f_h=1$ and $f_w=1$.  This filter, commonly utilized to manage model complexity, produces a single output element by performing a dot product between the weight tensor and a subvolume of an input activation, following a depth-first order. To accommodate both $1\times1$ convolutions~(and larger $f_h\times f_w$ filters), weights are usually stored and processed in a depth-first order.  Without loss of generality\footnote{Our observation is that similar classification accuracy tends to persist across different dimensional configurations as long as the total number of elements in the block is the same.} and to better reflects our target hardware architecture~(see Sec.~\ref{subsec:part}), we partition weights depth-wise into $[1,w]$ blocks, with the last block padded with zeros if necessary. It's worth noting that the $1\times1$ convolution example provided here serves as an illustration. The proposed method applies to larger convolutions (partitioned along depth) and matrix multiplication operations (partitioned along rows or columns) as well.
\subsection{Set Quantization Strategies}\label{subsec:setq}
%
\begin{figure}[h]
\centering
\includegraphics[width=1.0\columnwidth]{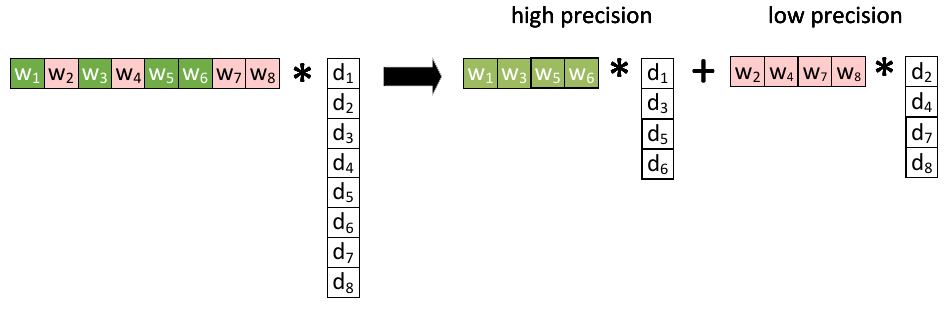}
\captionsetup{belowskip=0pt}
\caption{StruM with DLIQ, where $[l,w]=[1,8]$ and $p=0.5$.}
\label{fig:example_1_8}
\centering
\end{figure}
\begin{figure}[h]
\centering
\includegraphics[width=0.95\columnwidth]{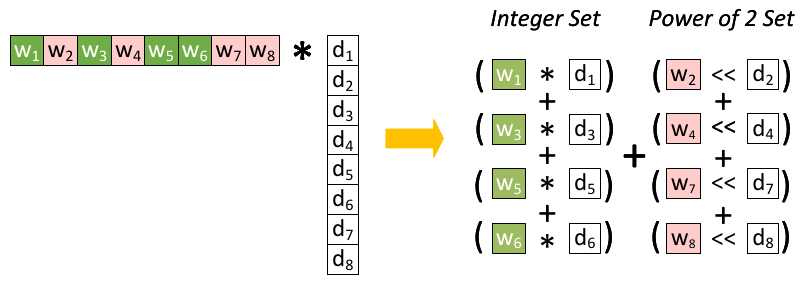}
\captionsetup{belowskip=0pt}
\caption{StruM with MIP2Q, where $[l,w]=[1,8]$ and $p=0.5$.}
\label{fig:example_1_8_shift}
\centering
\end{figure}

%
We now present three distinct strategies for partitioning each block into sets and quantizing them. Without loss of generalization, let's assume the initial weights are quantized to 8-bit (INT8) values. The first method, structured sparsity, is an existing approach used by NVIDIA. In this method, values within a block are divided based on their absolute magnitudes, with a split point determining which values receive a lower precision (0), while the rest remain unchanged. For equal-sized higher precision and lower precision sets (e.g., $p=0.5$), the split point is placed at the midpoint of the sorted list. This technique is natively implemented on NVIDIA's H100/A100 GPUs with parameters $l=1$, $w=4$, and $p=0.5$.

Overall, StruM employs the following two alternative methods to reduce hardware complexity.
\begin{enumerate}
    \item \textbf{Dual Level Integer Quantization~(DLIQ):} In this approach, like before, values are sorted by absolute magnitude, and those below a specified split point are quantized to a lower precision with $q$~bit (e.g., $q=4$ representing INT4), while values above the threshold remain unmodified. The parameter $p=0.5$ signifies equal higher precision (e.g., INT8) and lower precision (e.g., INT4) sets.

    \quad The main difference between this method and structured sparsity is the precision allowed for the lower precision set. 
    For example, in Fig.~\ref{fig:example_1_8}, given a $[1,8]$ block of values and $p=0.5$, structured sparsity assigns $50\%$ of the values to $0$ and $50\%$ of the values to 8 bit precision. The proposed method assigns $50\%$ of the values to a lower precision (e.g., INT4) and the other $50\%$ of the values to the original precision (8-bit in this case). Increasing the precision of the low precision reduces quantization error, ultimately improving network classification accuracy.
    
    \item \textbf{Mixed Integer and Power of Two Quantization~(MIP2Q):} This method minimizes error (e.g., L2 norm) by exploring partitions that maintain a fixed number of unmodified values while quantizing the rest to a power of 2. Let $x$ denote our original weights, $m$ denote a masking matrix that can only take values $\lbrace 0,1 \rbrace$ that has the same dimension as $x$, $\hat{x}$ denote our weights quantized to the nearest power of 2, $\bar{m}$ denote a masking matrix such that each element in $m$ is flipped~(i.e, $\bar{m}=1^{l\times w}-m$), and $\odot$ denote the Hadamard product. We determine:
    \begin{align*}
    \arg\,min_{m \in \lbrace 0,1 \rbrace ^ {l\times w}}{|| x - ((x\odot m ) + (\hat{x} \odot \bar{m}))||_2},\\
    \text{ subject to }{|m|_1 = p\times l\times w}
    \end{align*}
    Here, the search space is manageable, allowing for a comprehensive exploration of all possible partitions to find the one minimizing the error. 
    
    \quad Fig.~\ref{fig:example_1_8_shift} illustrates the operation of MIP2Q with a $[1,8]$ block of values and $p=0.5$. The primary distinction from DLIQ lies in the hardware required for implementation. DLIQ requires two types of multipliers: one for high precision and another for low precision. MIP2Q requires multipliers for one set and power-efficient arithmetic shifts for the other set, making it a more power-efficient choice.
    
    \quad It's possible to reduce the cost of arithmetic shifts. The range of all possible shifts (assuming INT8) can be represented using 4-bit values ($q=4$). This scenario corresponds to the case where $L=7$, corresponding to a shift range of $[-7,7]$. Complexity can be reduced by constraining the number of possible shifts. For example, setting $L=3$ confines the shift range to $[-3,3]$, which can be represented by a 3-bit value, offering a method to mitigate complexity. In general, the relationship between $q$ and $L$ is expressed as: $q=\left\lceil \log_2(L+1) \right\rceil+1$.
\end{enumerate}
%
%
\subsection{Runtime Processing}
The hardware accelerates computations for StruM through a multi-step process. Initially, weights are encoded, comprising both payload and mask header. The hardware efficiently utilizes the header information to perform dot products, enhancing computational efficiency.
\begin{figure}[h]
\centering
\includegraphics[width=1.0\columnwidth]{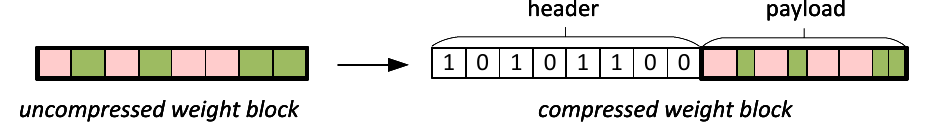}
\caption{Encoding uncompressed block into a compressed block}
\label{fig:packet}
\centering
\end{figure}
\subsubsection{Weight Encoding}
The hardware needs to identify which values within a block belong to which set to accelerate computation. We describe the encoding format used to store weights in a compressed format. In this compressed format, a mask bit in the header indicates whether the corresponding value within a block is stored in lower precision (denoted as $0$) or high precision (denoted as $1$). Fig~\ref{fig:packet} illustrates an example of how a $[1,8]$ uncompressed block with $p=0.5$ is stored as a compressed weight block. The mask bit is used to read the correct number of bits in the payload. In the example, the first bit of the mask header is $1$, indicating that the first weight value is in high precision. The second bit of the mask header is $0$, indicating that the second weight value is in low precision. To obtain the corresponding value, the hardware needs to read more bits from the payload for the first value compared to the second value.

The processing for the payload differs slightly for the two proposed methods. Consider high precision represented by 8-bit values and low precision represented by 4-bit values (denoted as $q=4$) in the context of our encoding scheme. In the case of DLIQ, when the first bit of the mask header is 1, an 8-bit value is retrieved from the payload. Conversely, if the first bit is 0, a 4-bit value is extracted from the payload.
In contrast, for MIP2Q, the mask bit in the header indicates whether the corresponding value within a block is stored as a power of 2 or an integer.  For example, when the first bit of the mask header is 1, an 8-bit value is retrieved from the payload. If the first bit is 0, we still read 4~bits from the payload, which we denote as k. However, the actual weight value is $2^k$, which, when multiplied, is simply an arithmetic shift by $k$.
\begin{figure}[h]
\centering
\includegraphics[width=0.90\columnwidth]{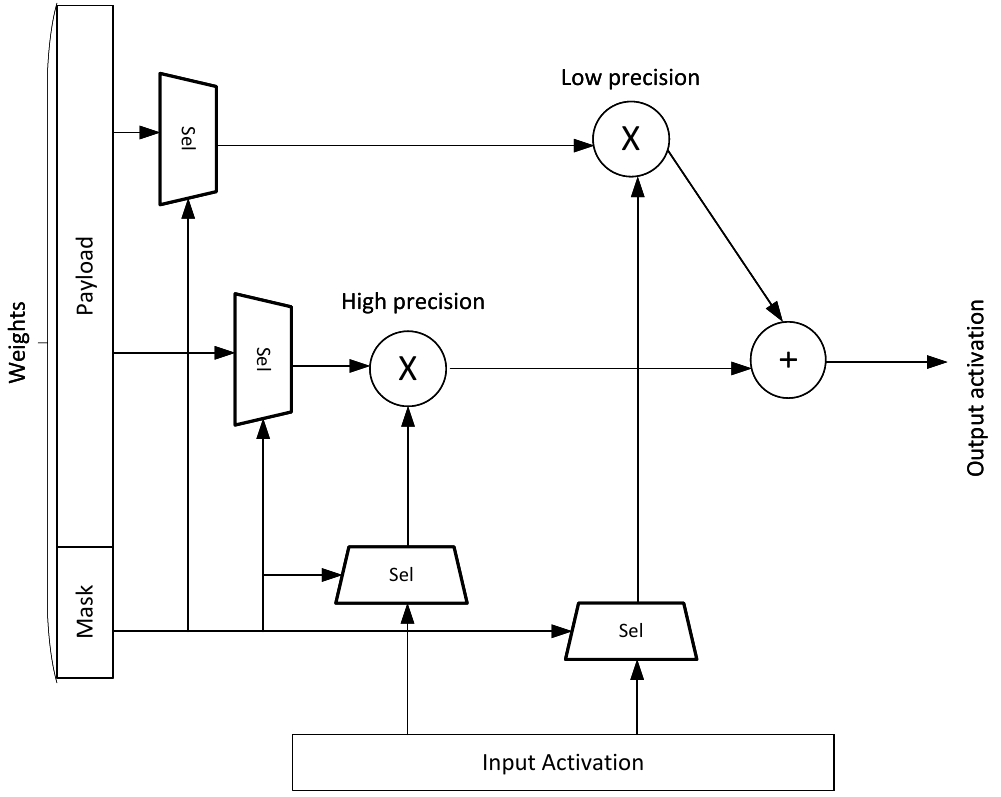}
\caption{Proposed StruM dot product multiplier}
\label{fig:dsp}
\centering
\end{figure}

The encoding format also reduces weight memory storage and bandwidth usage. Weight compression level ($r$) is defined as the ratio between compressed weight memory and uncompressed weight memory (compressed/uncompressed). In the previously mentioned weight encoding, the additional overhead arises from the mask header needed to manage the positions of low and high-precision elements within the $[l, w]$ block. Assuming an 8-bit representation for high precision and $q>1$, $r$ can be calculated as follows:
\begin{align} 
r = \frac{p(q-8)+9}{8} \label{eq:q>1}
\end{align}
Again note for MIP2Q, the relationship between $q$ and $L$ is expressed as: $q=\left\lceil \log_2(L+1) \right\rceil+1$.

We note that structured sparsity is a special case where there is no need to store lower precision values in the payload since these values are already known from the mask header. This also applies to the $q=1$ cases of DLIQ and MIP2Q. Assuming an 8-bit representation for high precision, $r$ can be computed as follows:
\begin{align}
r = \frac{9-8p}{8} \label{eq:q=1}
\end{align}
\subsubsection{Accelerated processing}
Finally, we explain how the hardware can leverage StruM to accelerate computations. Let's consider a dot product compute block capable of one dot product operation with vectors of length $m$ per clock cycle. Initially the dot product compute block is configured with $m$ INT8$\times$INT8 multipliers. We propose an alternative architecture that performs dot products using StruM. Fig.~\ref{fig:dsp} illustrates an example of a dot product computation with structured mixed precision. When $p=0.5$ and $[l,w]=[1,16]$, we still utilize $8$ INT8$\times$INT8 multipliers. However, the remaining $8$ multipliers become lower precision multipliers (e.g., INT4$\times$INT8) for DLIQ and arithmetic shifters for MIP2Q. This arrangement effectively provides 16 multipliers: 8 operating in high precision and 8 operating in low precision. During runtime, the mask header is utilized to partition the weight and activation values, directing them to the appropriate set of multipliers. Weight and activation values associated with a mask bit value of $1$ are routed to INT8$\times$INT8 multipliers, whereas those associated with a mask bit value of $0$ are directed to low precision multipliers. The partial accumulated products from each set of multipliers are then summed using an output accumulator.

\section{Microarchitecture Design}\label{sec_flexnn}

\begin{figure*}[ht]
\centering
\includegraphics[width=1.95\columnwidth]{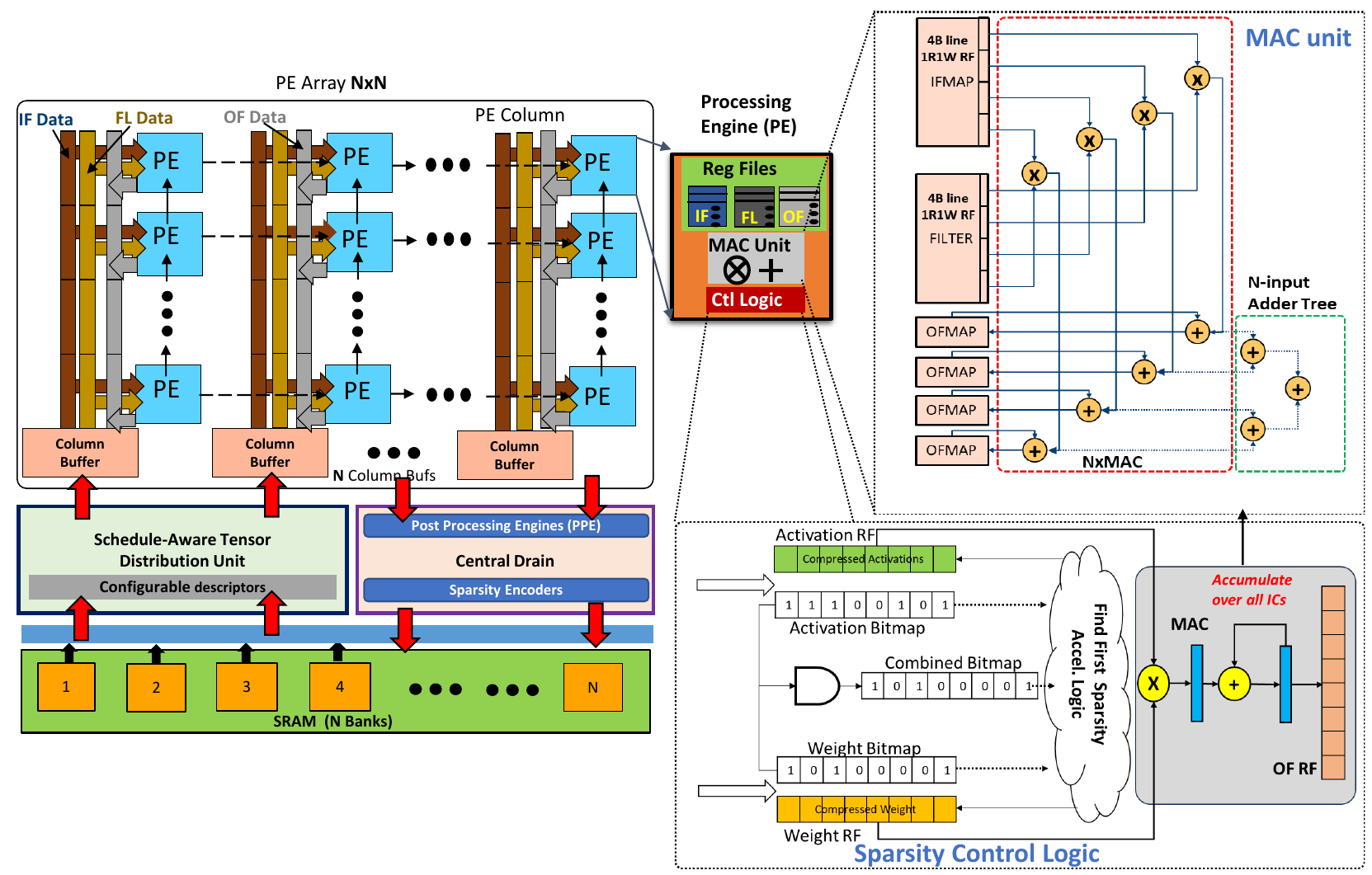}
\caption{FlexNN Array and PE Microarchitecture~\cite{flexnn}}
\label{fig:flexnn}
\centering
\end{figure*}


\begin{figure*}[ht]
\centering
\includegraphics[width=1.9\columnwidth]{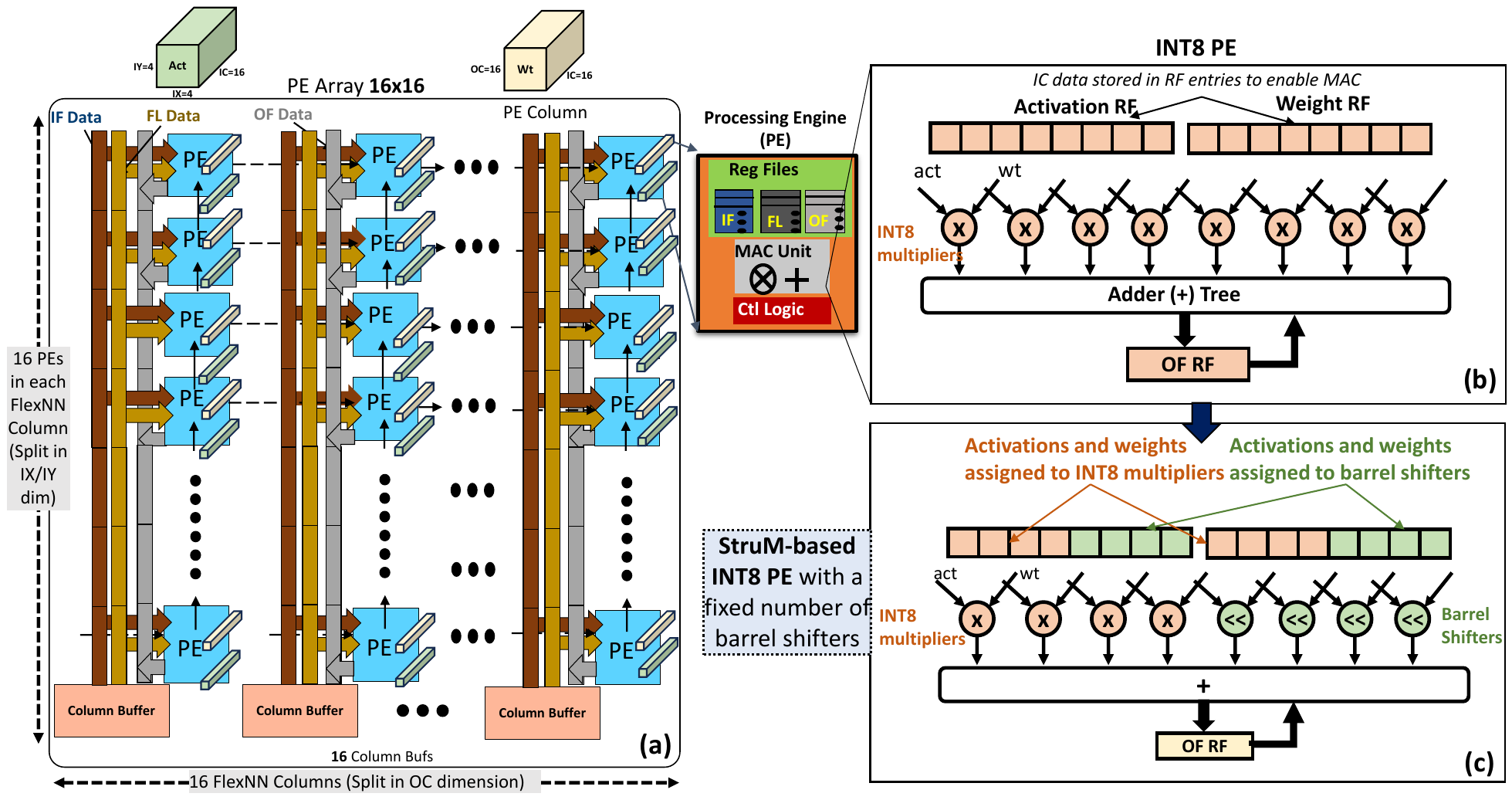}
\caption{StruM PE integrated to FlexNN accelerator~\cite{flexnn}}
\label{fig:top_level}
\centering
\end{figure*}

We leverage the FlexNN architecture~\cite{flexnn} to showcase the benefits of StruM. In this section, we first briefly introduce FlexNN, an existing flexible DNN accelerator hardware architecture described in detail in~\cite{flexnn}. FlexNN is a flexible schedule-aware DNN accelerator that can adapt its internal dataflow to the optimal schedule of each layer in DNNs. Subsequently, we describe our proposed StruM-based processing element (PE) variants within the DNN accelerator, aimed to reduce both complexity and power consumption. 

\subsection{High-Level Overview of FlexNN Accelerator Architecture}

Fig.~\ref{fig:top_level}(a, b) presents a high-level schematic of the FlexNN accelerator. The primary computing component within the DNN accelerator is the Multiply-and-Accumulate Compute (MAC) unit. A Processing Element (PE) consists of one or more MAC units. Typically, edge DNN accelerators use 8-bit integer~(INT8) MAC units for low-power execution. The PE depicted in Fig.~\ref{fig:top_level}(b) conducts an 8-wide dot product operation. This PE has eight INT8 multipliers, responsible for multiplying activations and weights in an 8-element array along the input channel (IC) dimension. Activation and weight values are respectively fetched from the local activation and weight \textit{Register Files~(RFs)}. After multiplication, the PE employs an adder tree, to sum up the eight resulting products before accumulating the result into the previous partial sum within a register (OF RF) within the PE. 

As shown in Fig.~\ref{fig:top_level}(a), the primary datapath of the DNN accelerator is the PE array, organized as a rectangular grid of processing elements (PEs) in a matrix of columns and rows. It is well-known that the MAC units within PEs contribute significantly to the overall PE (and thus the DNN accelerator’s) dynamic power consumption. This is because the MAC units are active nearly every clock cycle to maximize throughput or performance. Therefore, simplifying MAC operations has the potential to reduce the overall power consumption of the DNN accelerator notably. Note that we use the term data processing unit (or DPU) to refer to the complete DNN accelerator consisting of the PE array and the load and drain units. Fig.~\ref{fig:flexnn} also demonstrates the sparsity logic within the FlexNN accelerator. It depicts how the two-sided unstructured sparsity acceleration works within the FlexNN PE. The find-first logic identifies non-zero pairs of activations and weights using the incoming sparsity bitmaps and drives them through the multi-MAC-based FlexNN PE. In the case of NxMAC, the find-first logic finds the first N non-zero pairs of activations and weights in the RF and feeds to the MAC unit. The proposed StruM feature can be implemented on top of the existing sparsity logic.

\subsection{Accelerated PE with MIP2Q}
To simplify the MAC units within the PEs, we incorporated MIP2Q to reduce the complexity of the MAC units.\footnote{We opted for MIP2Q over DLIQ due to its demonstrated superior classification accuracy (refer to Sec.~\ref{sec:model_accuracy}).} The core concept involves implementing barrel shifters for low-precision multipliers~(see Fig.~\ref{fig:dsp}) to execute arithmetic shifts, thereby reducing overall complexity. In Fig.~\ref{fig:top_level}(c), we present our proposed INT8 PE, replacing $N$ instances of INT8-based multipliers with barrel shifters. Barrel shifters function as strength-reduced alternatives to INT8 multipliers, capable of executing a variable number of shift operations based on the inputs. For example, given inputs $A$ and $B>0$, barrel shifters can shift left the value of A by B positions, resulting in a product of $A \times 2^B$. Consequently, the scope of multiplication is limited to a subset of values representing powers of 2. In the proposed PE implementation, we assume $N = 4$, where we replace 4 out of 8 multiplication units with barrel shifters. We also explore two distinct MIP2Q PE variants. The first PE variant permits the full range of shifts~($L=7$). The second variant constrains the shift range to $[-5,5]$~(i.e. $L=5$). We opt for $L=5$ based on our findings in Sec.~\ref{sec:vary}, where we observed that the $L=5$ case closely approximates the accuracy of the unconstrained case. The $L=5$ variant allows for further hardware complexity reduction and power saving, as seen in Sec.~\ref{sec:power}.

\begin{figure*}[ht]
\centering
\includegraphics[width=1.9\columnwidth]{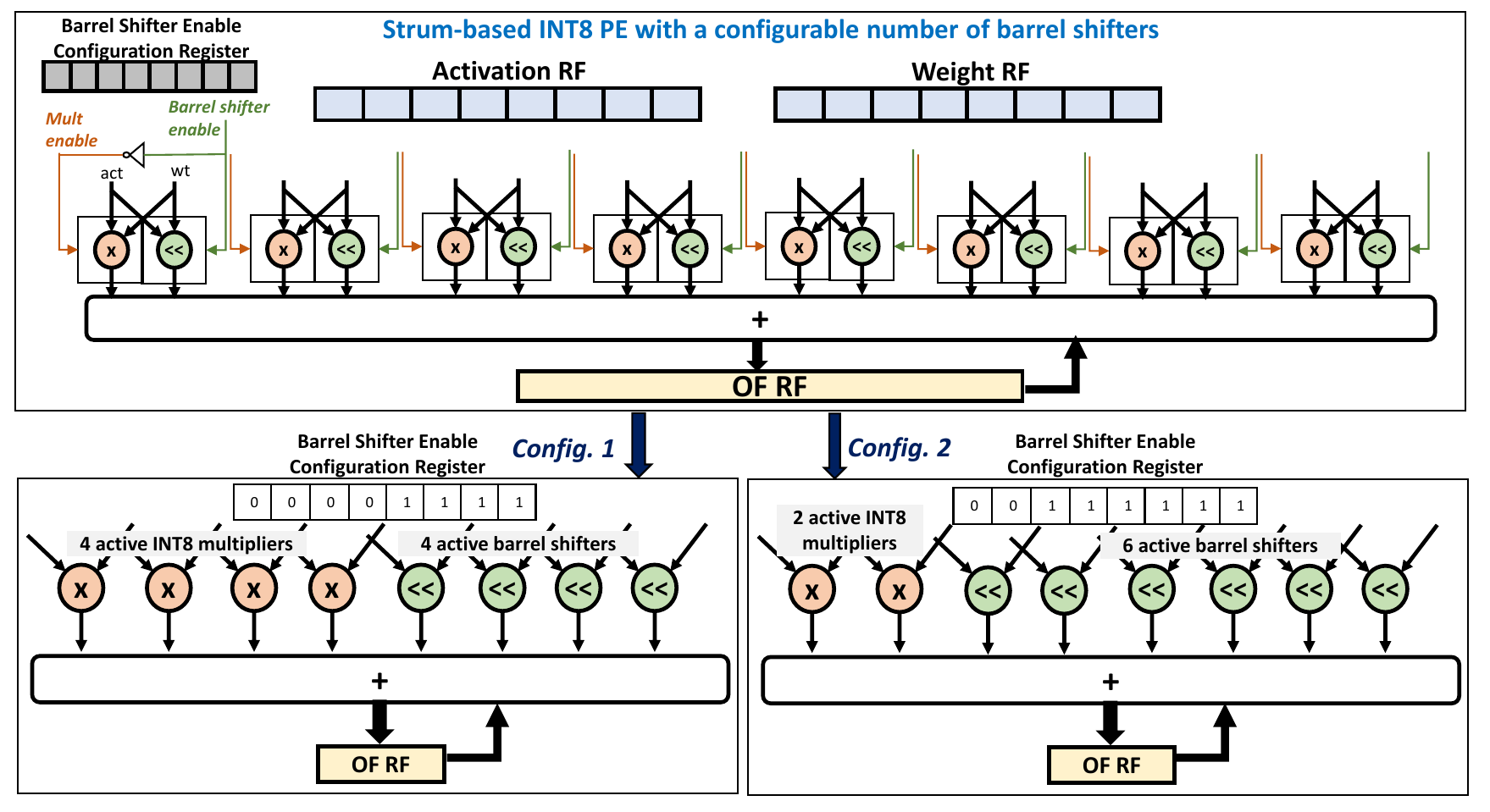}
\caption{StruM-based quality configurable PE with multipliers and barrel shifters}
\label{fig:barrshift}
\centering
\end{figure*}

We propose two different types of PE architecture for reducing the power of the PE. In the first type (Fig.~\ref{fig:top_level}(b)), we fix the number of multipliers to be replaced with barrel shifters during design time and the proportion of multipliers and shifters in each PE is fixed during runtime.  On the other hand, Fig.~\ref{fig:barrshift} shows a dynamically configurable PE that can decide during runtime the number of barrel shifters to be enabled in place of multipliers during convolution operation based on the specified quality degradation that the application can tolerate. This leads to disproportionate power savings at the DNN accelerator level in exchange for a small or negligible amount of quality loss.

Fig.~\ref{fig:barrshift} demonstrates two potential PE configurations. For the first one (Config. 1), we configure the number of active barrel shifters to be 4 whereas for the second one (Config. 2) the number of active barrel shifters is 6. The change in the PE architecture occurs through the programming of a barrel shifter enable configuration register that can be programmed via the compiler before each layer (or network) execution. Note that within the PE, the multiplier gets automatically disabled using clock gates when we enable the corresponding barrel shifter. For this particular case, since we instantiate extra barrel shifters in addition to multipliers there will be some extra area overhead compared to the previous case where we replace the multipliers altogether. The additional area overhead ensures we can configure the PEs based on the target quality or can even revert to the original quality if required in the worst case. Note that to limit the area overhead we integrate the barrel shifters in only a fixed number (N) of multipliers due to enforcing a fixed amount of structured mixed precision. For our experiments, we have included the barrel shifters in only half of the multipliers to limit the area overhead.

Note that for the case where we fix the number of multipliers and barrel shifters in the PE, we need to be backward compatible with the case where the layers require 8 multipliers per compute block instead of half of them (which is also the original requirement for the baseline INT8xINT8 operation in the FlexNN). For this mode, we assume that we will just use the 4 multipliers in a PE over 2 cycles, thus reducing the throughput by half. This should be fine as we assume the low precision case (with 4 multipliers and 4 shifters) to provide us with 2X acceleration at the layer level. So when we need 8 multiplications in a block of 8 we take 2 cycles to compute reducing the performance by 2X. This is another example where structured mixed precision can be really useful. Since we guarantee that in a block of compute there will be exactly 4 low precision values, we can guarantee that we will get a 2X acceleration in this low precision mode. For an unstructured case, we cannot guarantee such a performance uplift due to the imbalance of low precision values in different PEs.

\section{Experimental Methodology}\label{sec_exp_meth}
To assess the ImageNet classification accuracy of StruM, we first conducted static calibration using Graffitist~\cite{jain2019trained} to quantize both activations and weights to INT8 for all layers across all models. These INT8 quantized models served as our baseline before the application of any further optimization techniques. We then evaluated StruM for deep learning inference across various image classification neural networks by quantifying the top-1 accuracy on the ImageNet validation dataset.

We implemented StruM on top of the FlexNN architecture~\cite{flexnn} as stated before. The FlexNN PE with unstructured sparsity support is considered to be our baseline architecture. Note that since we reuse the sparsity bitmap as the precision bitmap for StruM, we operate the accelerator in the dense mode without any sparsity acceleration or sparsity compression for weights. However, theoretically, it is possible to enable both sparsity acceleration and mixed precision by utilizing two different bitmap encodings. However, this may increase the complexity of the design.

The FlexNN accelerator microarchitecture is written in a parameterized way. For StruM, we selected the baseline architecture to comprise of a unified tile of 256 PEs organized in a 16×16 grid (16 columns with each column having 16 individual PEs), featuring 8 MAC units (only 4 shown in Fig. 7 for brevity) within each PE, resulting in a total of 2048 MACs. This tile encompasses 1.5 MB of SRAM equipped with 32-byte read/write ports. The PE consists of 4x16 B IF Data RF Register File (RF), 4x16 B FL Data RF, and 16x4 B OF RF. In addition, each PE also consists of a 4x2B IF sparsity bitmap RF and 4x2B FL sparsity bitmap RF, which is one-eighth of the size of data RF as 1 bit in bitmap is used to represent 1 byte in data. Together, these RFs contribute to 208B RF per PE. The FlexNN PE can support both INT (I8, U8) and FP (FP16, BF16) modes of operation.

The [l.w] block values are closely aligned to the minimum granularity of computation within the FlexNN DPU. As stated earlier, we load the activations and weights at a minimum granularity of 16 ICs determined by the IF and FL RFs. Although the weights are loaded at a minimum granularity of 16 in the IC dimension, only a single OC is loaded into a FlexNN column as depicted in Fig.~\ref{fig:top_level}. This is because weights are usually broadcasted within a FlexNN column whereas activations are usually broadcasted across different FlexNN columns. Each FlexNN column has its weight (or OC) set. This enables us to load the different FlexNN columns with their different precision patterns independent to the other columns. Consequently, for our experiments we set the [l,w] = 1,16.

The StruM-based FlexNN PE and the DPU were first implemented in Chisel 3.0~\cite{chisel} (same as the original FlexNN). The generated SystemVerilog RTL was then simulated in Synopsys VCS. The RTL is subsequently mapped and synthesized on one of the industry's most advanced process technology nodes (based on a 3nm process). We used the Synopsys Fusion Compiler for synthesis and place \& route stages to prepare the final DPU layout. Post-synthesis, the layout was finalized to meet \textit{Quality of Result~(QoR)} requirements. The area, energy, and critical path delays were obtained post-layout after the route-optimization stage using a nominal voltage of 0.75V, operating at 1.8 GHz. To estimate the power consumption within the proposed FlexNN accelerator, we employed Synopsys Verdi to generate an activity file (Switching Activity Interchange Format: SAIF), using test benches for assistance. The accelerator netlist, coupled with the activity file, serves as input to Synopsys PrimeTimePX (PTPX), enabling power estimation at the gate level for both block and full-chip designs of the FlexNN accelerator.

Note that there is no compile time overhead for this feature. The only additional complexity added to the compiler is the identification of layers/groups/values that can perform well with shifters instead of multipliers. Since we perform these optimizations on the weights, this can be seen as a one-time effort spent during the encoding process.

\section{Experimental Results}\label{sec_Results_and_Discussions}
We have split the results section into two subsections. In the first subsection, we depict the model accuracy results for both DLIQ and MIP2Q-based structured mixed precision techniques. In the second part, we present the area and power savings acquired from the proposed StruM-based DNN accelerator.

\subsection{Model Accuracy Results}\label{sec:model_accuracy}
\noindent In this section, we evaluate StruM for deep learning inference across various image classification neural networks, focusing on ImageNet classification accuracy. We initiated our evaluation using the INT8 quantized model as our baseline. Subsequently, weights are partitioned into blocks and quantized for a second time, with a specified percentage, $p$, of values within each block quantized to a lower precision (e.g. $q=4$).  It is important to note that the achieved classification accuracies were obtained without the need for retraining or fine-tuning.
\begin{figure}
     \begin{subfigure}{0.24\textwidth}
         \centering
         \includegraphics[width=\textwidth, trim=6.5cm 9.25cm 6.5cm 10cm, clip]{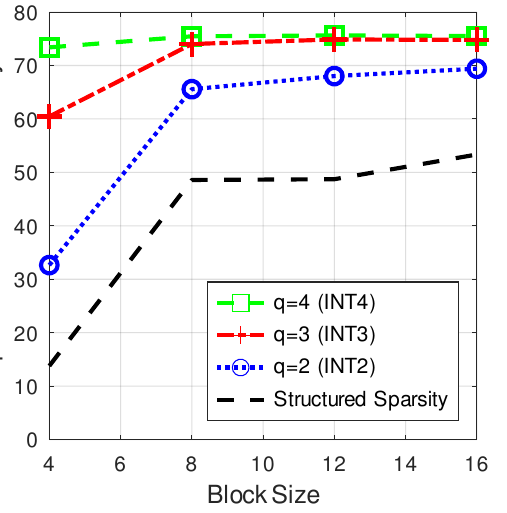}
         \caption{$p=0.5$}
         \label{fig:top1_p_tbq}
     \end{subfigure}
     \begin{subfigure}{0.24\textwidth}
         \centering
         \includegraphics[width=\textwidth, trim=6.5cm 9.25cm 6.5cm 10cm, clip]{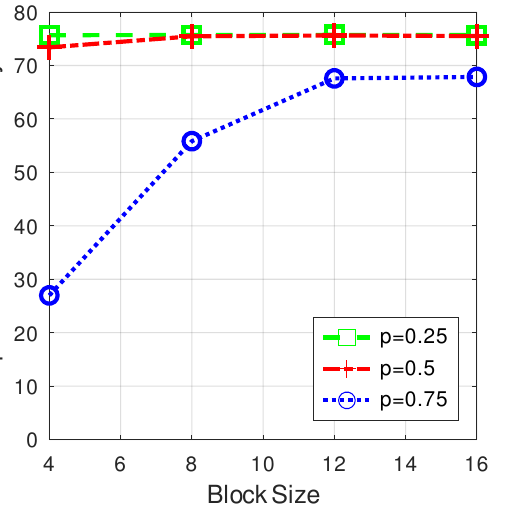}
         \caption{$q=4$~(INT4)}
         \label{fig:top1_q_tbq}
     \end{subfigure}
        \caption{DLIQ ImageNet Top-1 accuracy for Resnet-50 V1.5}
\end{figure}
\begin{figure}
     \begin{subfigure}{0.24\textwidth}
         \centering
         \includegraphics[width=\textwidth, trim=6.5cm 9.25cm 6.5cm 10cm, clip]{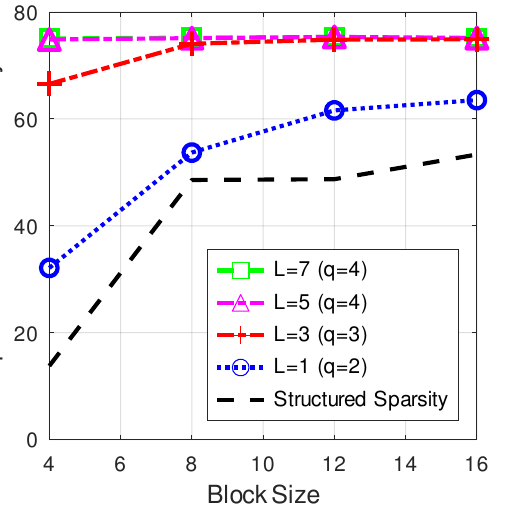}
         \caption{$p=0.5$}
         \label{fig:top1_p_mp2}
     \end{subfigure}
     \begin{subfigure}{0.24\textwidth}
         \centering
         \includegraphics[width=\textwidth, trim=6.5cm 9.25cm 6.5cm 10cm, clip]{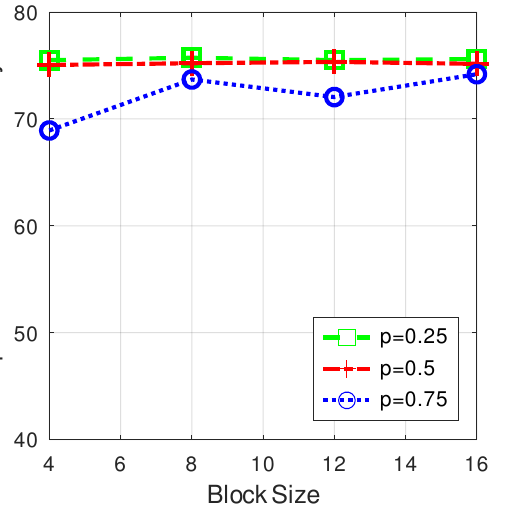}
         \caption{$q=4$}
         \label{fig:top1_q_mp2}
     \end{subfigure}
        \caption{MIP2Q ImageNet Top-1 accuracy for Resnet-50 V1.5}
\end{figure}

\subsubsection{Effects of Varying Parameters}\label{sec:vary}
We optimized the hardware implementation through ablation studies, varying the key parameters. Specifically, in Fig.~\ref{fig:top1_p_tbq} and Fig.~\ref{fig:top1_q_tbq}, we examine the impact of block size, $p$ and $q$ on the ImageNet classification accuracy of Resnet-50 v1.5 using DLIQ. Similarly, in Fig. \ref{fig:top1_p_mp2} and \ref{fig:top1_q_mp2}, we investigate the effects of block size, $p$, and $q$ on the ImageNet classification accuracy of Resnet-50 v1.5 utilizing MIP2Q.
The trends in DLIQ and MIP2Q are similar and are consistent with expectations:
\begin{enumerate}
\item Larger block sizes outperform smaller ones. This is expected as larger blocks make it easier to find small values or power-of-2 values within them. For instance, when the block size matches the filter size, the method reverts to the unstructured case, yielding the highest possible accuracy.
\item Smaller $p$ values  lead to superior accuracy compared to larger $p$ values. This is because smaller $p$ values result in less quantization error, quantizing fewer values within a block to a lower precision.
\item For DLIQ, larger $q$ values result in better accuracy than smaller $q$ values. This is as expected as larger bit-width assigned to lower precision values results in lower quantization error. Similarly, in MIP2Q, larger $L$ values enhance accuracy since they allow the representation of larger values (e.g. larger shifts) without clipping.
\end{enumerate}
Furthermore, it is crucial to note that the proposed methods both significantly outperform the structure sparsity case. Structured sparsity yields poor accuracy and necessitates retraining to approach the baseline accuracy levels. For MIP2Q, $L=5$ is an interesting case. The performance of $L=5$ is comparable to that of $L=7$. However, a low precision value for $L=5$ still needs 4 bits~$(q=4)$. Finally, MIP2Q demonstrates comparable performance to DLIQ for medium to large values of both $p$ and $q$. For small $q$ values, DLIQ outperforms MIP2Q, whereas in the context of large $p$ values, MIP2Q exhibits superior performance. 
\begin{figure}[h]
     \centering
     \begin{subfigure}[b]{0.24\textwidth}
         \centering
         {\includegraphics[width=\textwidth, trim=6.5cm 9.25cm 6.5cm 10cm, clip]{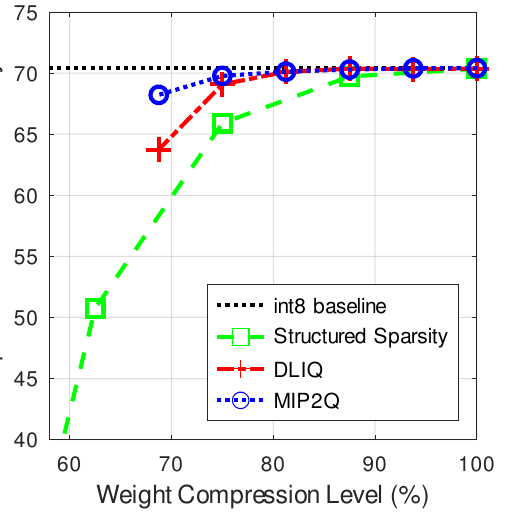}}
         \caption{VGG16}
         \label{fig:vgg16}
     \end{subfigure}
     \hfill
     \begin{subfigure}[b]{0.24\textwidth}
         \centering
         {\includegraphics[width=\textwidth, trim=6.5cm 9.25cm 6.5cm 10cm, clip]{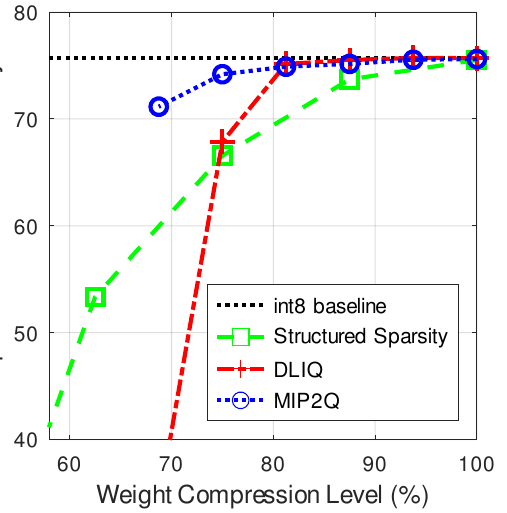}}
         \caption{Resnet-50 v1.5}
         \label{fig:resnet}
     \end{subfigure}
     \hfill
     \begin{subfigure}[b]{0.24\textwidth}
         \centering
         {\includegraphics[width=\textwidth, trim=6.5cm 9.25cm 6.5cm 10cm, clip]{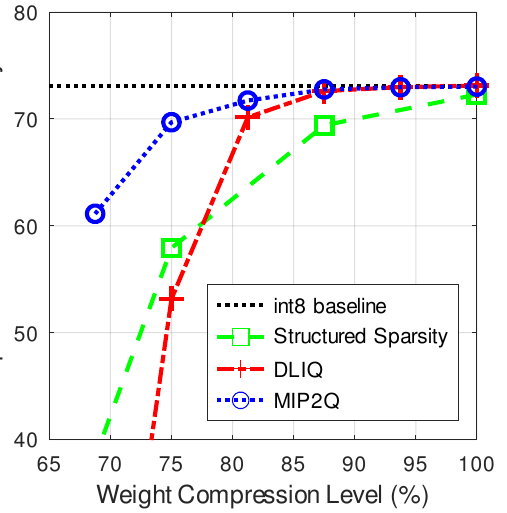}}
         \caption{Inception v2}
         \label{fig:inceptionv2}
     \end{subfigure}
     \hfill
          \begin{subfigure}[b]{0.24\textwidth}
         \centering
         {\includegraphics[width=\textwidth, trim=6.5cm 9.25cm 6.5cm 10cm, clip]{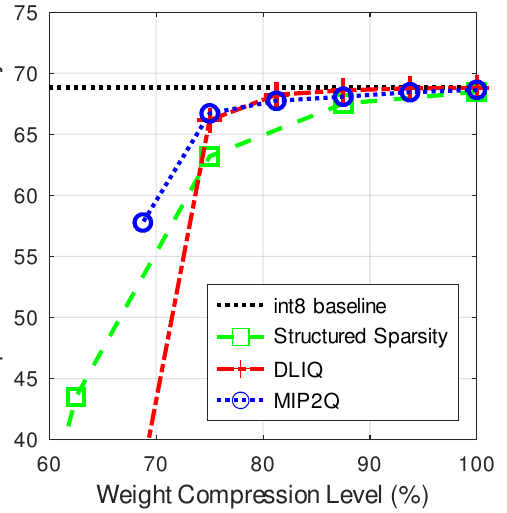}}
         \caption{Darknet-19}
         \label{fig:Darknet19}
     \end{subfigure}
        \caption{ImageNet Top-1 Classification Accuracy}
        \label{fig:three graphs}
\end{figure}

\begin{table*}[t]
\small
\centering
\begin{tabular}{|c|c|ccc|ccc|ccc|}
\hline 
\multirow{2}{*}{\textbf{Network}} & \multirow{2}{*}{\textbf{Baseline}} & \multicolumn{3}{c|}{\textbf{Structured Sparsity}} & \multicolumn{3}{c|}{\textbf{DLIQ}} & \multicolumn{3}{c|}{\textbf{MIP2Q}}\tabularnewline
\cline{3-11} \cline{4-11} \cline{5-11} \cline{6-11} \cline{7-11} \cline{8-11} \cline{9-11} \cline{10-11} \cline{11-11} 
 &  & \textbf{$p=0.25$} & \textbf{$p=0.50$} & \textbf{$p=0.75$} & \textbf{$p=0.25$} & \textbf{$p=0.50$} & \textbf{$p=0.75$} & \textbf{$p=0.25$} & \textbf{$p=0.50$} & \textbf{$p=0.75$}\tabularnewline
\hline 
\textbf{VGG16} & 70.4 & 69.7 & \textbf{50.7} & \textbf{0.2} & 70.4 & 70.4 & \textbf{69.1} & 70.4 & 70.3 & 69.8\tabularnewline
\textbf{VGG19} & 70.4 & 69.9 & \textbf{45.5} & \textbf{0.2} & 70.4 & 70.3 & \textbf{69.0} & 70.5 & 70.4 & 70.1\tabularnewline
\textbf{Resnet-50 v1.5} & 75.7 & \textbf{73.7} & \textbf{53.3} & \textbf{0.4} & 75.7 & 75.5 & \textbf{67.9} & 75.7 & 75.2 & \textbf{74.2}\tabularnewline
\textbf{Resnet-101} & 74.8 & \textbf{72.6} & \textbf{69.0} & \textbf{0.5} & 74.7 & 74.7 & \textbf{72.6} & 74.7 & 74.4 & \textbf{72.5}\tabularnewline
\textbf{Resnet-152} & 76.2 & 75.6 & \textbf{70.6} & \textbf{0.9} & 76.2 & 76.2 & \textbf{74.4} & 76.2 & 75.8 & \textbf{75.0}\tabularnewline
\textbf{Inception V1} & 68.6 & \textbf{63.2} & \textbf{19.6} & \textbf{0.1} & 68.6 & 68.1 & \textbf{58.7} & 68.6 & 68.0 & \textbf{63.7}\tabularnewline
\textbf{Inception V2} & 73.1 & \textbf{69.4} & \textbf{18.7} & \textbf{0.1} & 73.1 & 72.6 & \textbf{53.1} & 73.1 & 72.7 & \textbf{69.7}\tabularnewline
\textbf{Inception V3} & 76.8 & \textbf{72.8} & \textbf{14.4} & \textbf{0.1} & 76.9 & 76.6 & \textbf{70.9} & 76.8 & 76.5 & \textbf{70.3}\tabularnewline
\textbf{Inception V4} & 79.5 & \textbf{78.2} & \textbf{36.1} & \textbf{0.1} & 79.4 & 79.4 & \textbf{70.9} & 79.4 & 79.1 & \textbf{72.3}\tabularnewline
\textbf{Darknet-19} & 68.8 & \textbf{67.5} & \textbf{43.5} & \textbf{0.1} & 68.8 & 68.6 & \textbf{66.2} & 68.8 & \textbf{68.1} & \textbf{66.7}\tabularnewline
\hline 
\end{tabular}
\caption{Top-1 ImageNet classification error comparison: For all cases, the block size is $[1,16]$. The baseline utilizes INT8 quantization for both activations and weights.   \textbf{Bold} values indicate cases where the relative Top-1 classification error is more than $1\%$ above the baseline.}
\label{tbl:top1_combined}
\end{table*}

\subsubsection{Classification Accuracy Across Networks}\label{sec:class_results}
In Table~\ref{tbl:top1_combined}, we present the Top-1 classification accuracy of the proposed methods across various networks on the ImageNet dataset. We employed a block size of $[l,w]=[1,16]$ for both methods and varied $p$. For consistency, we set $q=4$ for both DLIQ and MIP2Q. Remarkably, we observed minimal relative accuracy loss (less than $1\%$) compared to the INT8 baseline across all networks for both methods at $p=0.25$ and $p=0.5$. As discussed in Sec.~\ref{sec:vary}, our observations reveal that for $p=0.75$, MIP2Q consistently outperforms DLIQ. However, it is crucial to note that in this particular scenario, we observed a greater than $1\%$ relative accuracy loss from both methods. Finally, across all cases, DLIQ and MIP2Q consistently achieved better classification accuracy than structured sparsity.

However, for the same $q$, structured sparsity requires a smaller weight storage requirement compared to DLIQ and MIP2Q (refer to Eq.\ref{eq:q>1} and Eq.\ref{eq:q=1}). In Fig.~\ref{fig:three graphs}, we compare Top-1 classification accuracy against weight compression level~($r$) for all three methods. The absolute accuracy difference is network dependent. We observe that for large $r$ values, both DLIQ and MIP2Q outperform sparsity. Interestingly, structured sparsity outperforms DLIQ for smaller $r$ values. However, MIP2Q outperforms both structured sparsity and DLIQ for smaller $r$ values. Given the demonstrated stability in accuracy across different parameter combinations and achieving similar performance to DLIQ, we selected MIP2Q for our hardware implementation.

\subsection{DPU-level Area and Power Savings}\label{sec:power}
\begin{figure}[h]
\centering
\includegraphics[width=\columnwidth]{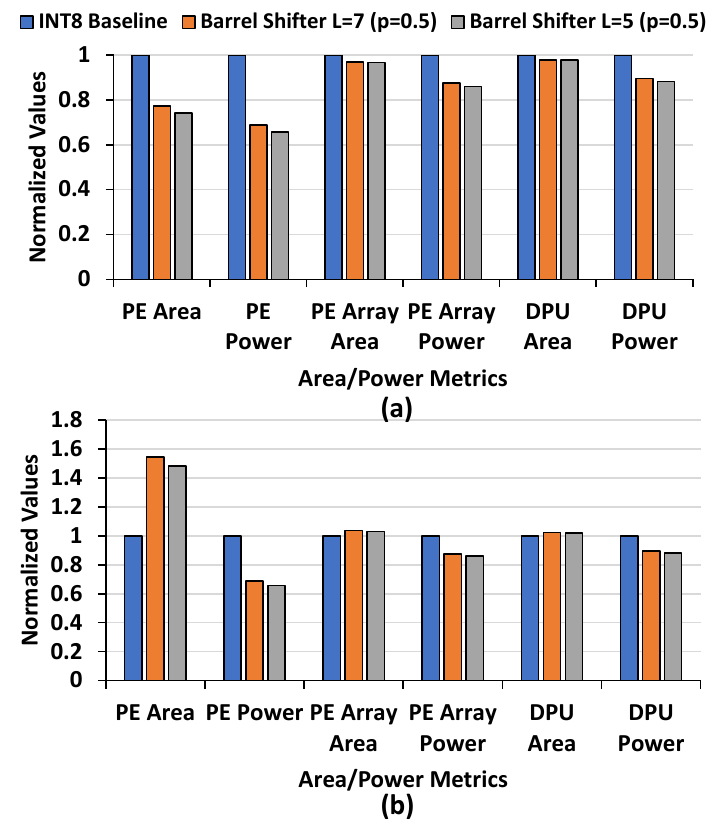}
\caption{DPU-level, PE Array-level, and PE-level Area/Power Metrics for a fixed number of barrel shifters compared to the baseline with only multipliers. (a) Permanent replacement of multipliers with barrel shifters, (b) Configurable PE with both multipliers and barrel shifters.}
\label{fig:shift_results}
\centering
\end{figure} 


For most client- or edge-based AI accelerators, TOPS/W and TOPS/mm2 are the two most important evaluation/comparison metrics.  With the proposed StruM technique, we can significantly improve both TOPS/W and TOPS/mm2. This can be achieved if we permanently replace some of the multipliers with barrel shifters. Otherwise, if our target is just to improve the TOPS/W, we can instantiate the barrel shifters on top of the multipliers. Although the latter would lead to some additional area overhead, it can still result in the same TOPS/W improvement as before, as the multipliers can be gated when we use the shifters. However, in exchange for the area penalty, we get the added advantage of defaulting back to the multipliers when the target accuracy cannot be met. The current critical path of the FlexNN design resides within the floating-point MAC of the FlexNN PE as the latter also supports inference in FP16/BF16 formats. The addition of the barrel shifter does not impact the critical path in any way and the timing bottleneck remains within the FPMAC.

Fig.~\ref{fig:shift_results} depict the area and power savings resulting from statically and dynamically configured barrel shifter-based PEs, respectively for $L=7$ and $L=5$ cases where $p=0.5$ (that replaces half of the INT8 multipliers with barrel shifters). Note that our baseline numbers are based on the FlexNN architecture~\cite{flexnn}. We present the area and power-saving results of the two proposed MIP2Q PE variants: one utilizing a full-range barrel shifter with $L=7$ and the other employing a reduced-range barrel shifter with $L=5$. When analyzing PE area/power in isolation, both variants demonstrate over $23-26\%$ and $31-34\%$ in area and power savings, respectively. Specifically, the $L=7$ case yields approximately $23\%$ area savings and around $31\%$ power savings at the PE level. The $L=5$ variant results in a slightly higher degree of area ($26\%$) and power ($34\%$) savings. However, when considering the PE array or DPU (accelerator)  as a whole, significant overhead (such as the register file) imposes limitations on the relative area savings for both variants. Nevertheless, MAC operations make a substantial contribution to the overall power consumption of the PE Array and DNN. Consequently, both PE variants exhibit over $10-12\%$ power savings when considering the entirety of the PE array or DPU. Similar to the previous findings, the $L=5$ variant slightly reduces PE array/DPU power compared to the $L=7$ case but it also comes at a cost of lower accuracy. The statically configured shift-based PE results in $2-3\%$ area benefits over the multiplier-based baseline at the DPU level. On the other hand, the dynamically configurable barrel shifter-based PE has a small area overhead of $3\%$ at the DPU level.


From Sec.~\ref{sec_Results_and_Discussions}, we see $L=7$ results in an almost negligible loss in accuracy compared to the INT8 multiplier-only baseline. We see that $L=5$ is a viable alternative, as it attains comparable accuracy to $L=7$ while offering additional power reduction.  This highlights the meaningful trade-off achieved by both cases, enhancing power efficiency without compromising classification accuracy.

\section{Conclusion}\label{sec_Conclusion_and_Future_Work}
This paper introduces StruM, a novel approach aimed at enhancing inference efficiency by partitioning weight blocks into two distinct sets with varying precisions. This strategy involves processing these sets at different precisions to maximize computational efficiency. Our investigation involved comparing two alternative quantization strategies, DLIQ and MIP2Q, revealing similar performance between the two, with MIP2Q demonstrating superior performance across various parameters such as block sizes and $p$. Notably, both approaches closely align with the INT8 baseline. To demonstrate StruM's compatibility with a hardware implementation, we enhanced the PE within our DPU using MIP2Q. This involved substituting a subset of multipliers with hardware-efficient barrel shifters. Our findings indicate substantial savings at the PE level, roughly around $23-26\%$ reduction in area and $31-34\%$ power consumption. At the DPU level, we achieved a notable $10\%$ reduction in power consumption. At present, $p$, the percentage representing lower precision values for layers remains constant. In future work, we aim to explore methods for dynamically adjusting $p$ on a per-layer basis. Additionally, we plan to investigate dynamically configurable PEs that can determine, at runtime, the optimal number of values to assign to low-precision multipliers during convolution operations, which may allow further reductions in power consumption.


\bibliographystyle{./bibliography/IEEEtran}
\bibliography{./bibliography/IEEEabrv,./bibliography/DAC}

\begin{thebibliography}{10}
\providecommand{\url}[1]{#1}
\csname url@samestyle\endcsname
\providecommand{\newblock}{\relax}
\providecommand{\bibinfo}[2]{#2}
\providecommand{\BIBentrySTDinterwordspacing}{\spaceskip=0pt\relax}
\providecommand{\BIBentryALTinterwordstretchfactor}{4}
\providecommand{\BIBentryALTinterwordspacing}{\spaceskip=\fontdimen2\font plus
\BIBentryALTinterwordstretchfactor\fontdimen3\font minus \fontdimen4\font\relax}
\providecommand{\BIBforeignlanguage}[2]{{%
\expandafter\ifx\csname l@#1\endcsname\relax
\typeout{** WARNING: IEEEtran.bst: No hyphenation pattern has been}%
\typeout{** loaded for the language `#1'. Using the pattern for}%
\typeout{** the default language instead.}%
\else
\language=\csname l@#1\endcsname
\fi
#2}}
\providecommand{\BIBdecl}{\relax}
\BIBdecl

\bibitem{flexnn}
A.~Raha, D.~A. Mathaikutty, S.~K. Ghosh, and S.~Kundu, ``{FlexNN}: A dataflow-aware flexible deep learning accelerator for energy-efficient edge devices,'' \emph{arXiv preprint arXiv:2403.09026}, 2024.

\bibitem{szegedy2015going}
C.~Szegedy, W.~Liu, Y.~Jia, P.~Sermanet, S.~Reed, D.~Anguelov, D.~Erhan, V.~Vanhoucke, and A.~Rabinovich, ``Going deeper with convolutions,'' in \emph{Proc. of CVPR}, 2015, pp. 1--9.

\bibitem{iandola2016squeezenet}
F.~N. Iandola, S.~Han, M.~W. Moskewicz, K.~Ashraf, W.~J. Dally, and K.~Keutzer, ``{SqueezeNet}: {AlexNet}-level accuracy with 50x fewer parameters and $<0.5$~{MB} model size,'' \emph{arXiv preprint arXiv:1602.07360}, 2016.

\bibitem{howard2017mobilenets}
A.~G. Howard, M.~Zhu, B.~Chen, D.~Kalenichenko, W.~Wang, T.~Weyand, M.~Andreetto, and H.~Adam, ``{MobileNets}: Efficient convolutional neural networks for mobile vision applications,'' \emph{arXiv preprint arXiv:1704.04861}, 2017.

\bibitem{tan2021efficientnetv2}
M.~Tan and Q.~V. Le, ``{EfficientNetV2}: Smaller models and faster training,'' \emph{arXiv preprint arXiv:2104.00298}, 2021.

\bibitem{zhang2018shufflenet}
X.~Zhang, X.~Zhou, M.~Lin, and J.~Sun, ``Shufflenet: An extremely efficient convolutional neural network for mobile devices,'' in \emph{Proceedings of the IEEE conference on computer vision and pattern recognition}, 2018, pp. 6848--6856.

\bibitem{chollet2017xception}
F.~Chollet, ``Xception: Deep learning with depthwise separable convolutions,'' in \emph{Proceedings of the IEEE conference on computer vision and pattern recognition}, 2017, pp. 1251--1258.

\bibitem{zoph2017neural}
\BIBentryALTinterwordspacing
B.~Zoph and Q.~Le, ``Neural architecture search with reinforcement learning,'' in \emph{International Conference on Learning Representations}, 2017. [Online]. Available: \url{https://openreview.net/forum?id=r1Ue8Hcxg}
\BIBentrySTDinterwordspacing

\bibitem{cai2018proxylessnas}
H.~Cai, L.~Zhu, and S.~Han, ``Proxyless{NAS}: Direct neural architecture search on target task and hardware,'' in \emph{Proc. of ICLR}, 2019.

\bibitem{tan2019mnasnet}
M.~Tan, B.~Chen, R.~Pang, V.~Vasudevan, M.~Sandler, A.~Howard, and Q.~V. Le, ``{MnasNet}: Platform-aware neural architecture search for mobile,'' in \emph{Proceedings of the IEEE/CVF conference on computer vision and pattern recognition}, 2019, pp. 2820--2828.

\bibitem{zhou2016dorefa}
S.~Zhou, Y.~Wu, Z.~Ni, X.~Zhou, H.~Wen, and Y.~Zou, ``{DoReFa-Net}: Training low bitwidth convolutional neural networks with low bitwidth gradients,'' \emph{arXiv preprint arXiv:1606.06160}, 2016.

\bibitem{hubara2016binarized}
I.~Hubara, M.~Courbariaux, D.~Soudry, R.~El-Yaniv, and Y.~Bengio, ``Binarized neural networks,'' \emph{Advances in neural information processing systems}, vol.~29, 2016.

\bibitem{umuroglu2017finn}
Y.~Umuroglu, N.~J. Fraser, G.~Gambardella, M.~Blott, P.~Leong, M.~Jahre, and K.~Vissers, ``{FINN}: A framework for fast, scalable binarized neural network inference,'' in \emph{Proc. of FPGA}, 2017, pp. 65--74.

\bibitem{gong2018highly}
J.~Gong, H.~Shen, G.~Zhang, X.~Liu, S.~Li, G.~Jin, N.~Maheshwari, E.~Fomenko, and E.~Segal, ``Highly efficient 8-bit low precision inference of convolutional neural networks with {IntelCaffe},'' in \emph{Proc. of Reproducible Quality-Efficient Systems Tournament on Co-designing Pareto-efficient Deep Learning}, 2018, p.~1.

\bibitem{jain2019trained}
S.~R. Jain, A.~Gural, M.~Wu, and C.~H. Dick, ``Trained quantization thresholds for accurate and efficient fixed-point inference of deep neural networks,'' 2019.

\bibitem{zhu2020xor}
S.~Zhu, L.~H. Duong, and W.~Liu, ``{XOR-Net}: An efficient computation pipeline for binary neural network inference on edge devices,'' in \emph{2020 IEEE 26th international conference on parallel and distributed systems (ICPADS)}.\hskip 1em plus 0.5em minus 0.4em\relax IEEE, 2020, pp. 124--131.

\bibitem{han2015deep}
S.~Han, H.~Mao, and W.~J. Dally, ``Deep compression: Compressing deep neural networks with pruning, trained quantization and huffman coding,'' \emph{arXiv preprint arXiv:1510.00149}, 2015.

\bibitem{han2015learning}
S.~Han, J.~Pool, J.~Tran, and W.~Dally, ``Learning both weights and connections for efficient neural network,'' \emph{Advances in neural information processing systems}, vol.~28, 2015.

\bibitem{li2016pruning}
H.~Li, A.~Kadav, I.~Durdanovic, H.~Samet, and H.~P. Graf, ``Pruning filters for efficient {ConvNets},'' \emph{arXiv preprint arXiv:1608.08710}, 2016.

\bibitem{narang2017block}
S.~Narang, E.~Undersander, and G.~Diamos, ``Block-sparse recurrent neural networks,'' \emph{arXiv preprint arXiv:1711.02782}, 2017.

\bibitem{gray2017gpu}
S.~Gray, A.~Radford, and D.~P. Kingma, ``{GPU} kernels for block-sparse weights,'' \emph{arXiv preprint arXiv:1711.09224}, vol.~3, 2017.

\bibitem{mishra2021accelerating}
A.~Mishra, J.~A. Latorre, J.~Pool, D.~Stosic, D.~Stosic, G.~Venkatesh, C.~Yu, and P.~Micikevicius, ``Accelerating sparse deep neural networks,'' \emph{arXiv preprint arXiv:2104.08378}, 2021.

\bibitem{cottier2024rising}
B.~Cottier, R.~Rahman, L.~Fattorini, N.~Maslej, and D.~Owen, ``The rising costs of training frontier {AI} models,'' \emph{arXiv preprint arXiv:2405.21015}, 2024.

\bibitem{jia2022xvdpu}
X.~Jia, Y.~Zhang, G.~Liu, X.~Yang, T.~Zhang, J.~Zheng, D.~Xu, Z.~Liu, M.~Liu, X.~Yan \emph{et~al.}, ``{XVDPU}: A high performance cnn accelerator on versal platform powered by ai engine,'' \emph{ACM Transactions on Reconfigurable Technology and Systems}, 2022.

\bibitem{choquette2023nvidia}
J.~Choquette, ``Nvidia hopper h100 gpu: Scaling performance,'' \emph{IEEE Micro}, 2023.

\bibitem{luo2017thinet}
J.-H. Luo, J.~Wu, and W.~Lin, ``{ThiNet}: A filter level pruning method for deep neural network compression,'' in \emph{Proceedings of the IEEE international conference on computer vision}, 2017, pp. 5058--5066.

\bibitem{siswanto2021reconciling}
A.~Siswanto, J.~Frankle, and M.~Carbin, ``Reconciling sparse and structured pruning: A scientific study of block sparsity,'' in \emph{Workshop paper at the 9th International Conference on Learning Representations (ICLR 2021)}, 2021.

\bibitem{pool2020accelerating}
J.~Pool, ``Accelerating sparsity in the {NVIDIA} {Ampere} architecture,'' \emph{GTC 2020}, 2020.

\bibitem{tung2018deep}
F.~Tung and G.~Mori, ``Deep neural network compression by in-parallel pruning-quantization,'' \emph{IEEE transactions on pattern analysis and machine intelligence}, vol.~42, no.~3, pp. 568--579, 2018.

\bibitem{lee2017lognet}
E.~H. Lee, D.~Miyashita, E.~Chai, B.~Murmann, and S.~S. Wong, ``{LogNet}: Energy-efficient neural networks using logarithmic computation,'' in \emph{2017 IEEE International Conference on Acoustics, Speech and Signal Processing (ICASSP)}.\hskip 1em plus 0.5em minus 0.4em\relax IEEE, 2017, pp. 5900--5904.

\bibitem{elhoushi2021deepshift}
M.~Elhoushi, Z.~Chen, F.~Shafiq, Y.~H. Tian, and J.~Y. Li, ``Deepshift: Towards multiplication-less neural networks,'' in \emph{Proceedings of the IEEE/CVF conference on computer vision and pattern recognition}, 2021, pp. 2359--2368.

\bibitem{DRQ}
Z.~Song, B.~Fu, F.~Wu, Z.~Jiang, L.~Jiang, N.~Jing, and X.~Liang, ``{DRQ}: Dynamic region-based quantization for deep neural network acceleration,'' in \emph{2020 ACM/IEEE 47th Annual International Symposium on Computer Architecture (ISCA)}.\hskip 1em plus 0.5em minus 0.4em\relax IEEE, 2020, pp. 1010--1021.

\bibitem{HAQ}
K.~Wang, Z.~Liu, Y.~Lin, J.~Lin, and S.~Han, ``{HAQ}: Hardware-aware automated quantization with mixed precision,'' in \emph{Proceedings of the IEEE/CVF conference on computer vision and pattern recognition}, 2019, pp. 8612--8620.

\bibitem{chisel}
J.~Bachrach, H.~Vo, B.~Richards, Y.~Lee, A.~Waterman, R.~Avi\v{z}ienis, J.~Wawrzynek, and K.~Asanovi\'{c}, ``Chisel: constructing hardware in a scala embedded language,'' in \emph{Proceedings of the Design Automation Conference (DAC)}, 2012, p. 1216–1225.

\end{thebibliography}

\end{document}